\documentclass[reprint,superscriptaddress,longbibliography,amsmath,amssymb,aps,prb]{revtex4-2}
\usepackage[english]{babel}
\usepackage[T1]{fontenc}
\usepackage{lmodern}
\usepackage{amsmath}
\usepackage{amssymb}
\usepackage{amsfonts}
\usepackage{amsthm}
\usepackage{graphicx}    
\usepackage[dvipsnames]{xcolor}
\usepackage{hyperref}
\hypersetup{   
        unicode=true,           
        pdftoolbar=true,        
        pdfmenubar=true,        
        pdffitwindow=false,     
        pdfstartview={FitH},    
        pdfauthor={Adriano Amaricci},
        pdftitle={Entanglement-informed distributed wavefunction approach to scalable
        quantum many-body systems},
        pdfkeywords={quantum many-body systems, strongly correlated electrons, quantum spin systems, computational methods, entanglement},
        pdfnewwindow=true,      
        colorlinks=true,        
        linkcolor=MidnightBlue,        
        citecolor=Maroon,        
        filecolor=Brown,        
        urlcolor=Maroon          
}

\newcommand{\equ}[1]
{Eq.~(\ref{#1})}

\newcommand{\figu}[1]
{Fig.~\ref{#1}}

\newcommand{\ket}[1]
{|#1\rangle}

\newcommand{\bra}[1]
{\langle #1|}

\def\a{\alpha}

\def\CC{{\cal C}}\def\HH{{\cal H}}
 
 \def\OO{{\cal O}}
\def\QQ{{\cal Q}}

 \def\SS{{\cal S}}

 \def\CCC{\mathbb{C}}

\def\=={\equiv}

\def\=={\equiv}
 
  \def\Tr{{\rm Tr}\,}

\def\qa{{\bf q}}

\usepackage{bbold}
\def\11{\mathbb{1}}
\def\00{\mathbf{0}}

\makeatletter
\newcommand\footnoteref[1]{\protected@xdef\@thefnmark{\ref{#1}}\@footnotemark}
\makeatother

\begin{document}

\author{A.~Amaricci}
\affiliation{CNR-IOM, Istituto Officina dei Materiali, Consiglio Nazionale delle
    Ricerche, Via Bonomea 265, 34136 Trieste, Italy}

\title{Entanglement-informed distributed wavefunction approach to scalable
  quantum many-body systems}
\begin{abstract}
  We show that the entanglement structure of quantum many-body states
  defines a natural and optimal distributed representation for their
  simulation. An arbitrary entanglement cut induces a bipartite
  decomposition of the wavefunction, mapping its distribution onto that
  of the entanglement spectrum. In this representation the Hamiltonian
  application, the core of Krylov-subspace methods, reduces to local
  contractions and communication-optimal operations. Using
  benchmarks from different methods and models, we demonstrate
  near-linear scaling for sufficiently large systems and identify
  entanglement spectrum fragmentation as a key factor controlling 
  computational cost.
  This establishes entanglement as an
  organizing principle and unified, method-independent, route for
  scaling up quantum many-body simulations.
\end{abstract}

\maketitle

%
%
Quantum many-body systems (QMBs) lie at the core of modern physics, from
condensed matter to quantum chemistry and quantum information
science~\cite{Kotliar2004PT,Amico2008RMP,Bradlyn2017N,Tokura2017NP,Fang2026A,Kuramoto2020,Cui2022S,Huang2022S,Wu2024S,Bellomia2024PR,Tafuri2026}.
Across these contexts the key challenge is nearly identical: investigate
quantum phenomena arising from interactions between many degrees of
freedom and transcending the properties of individual particles, e.g.
superconductivity~\cite{Gunnarsson1997RMP,Capone2002S,Lee2006RMP,Anderson2007S,Capone2009RMP,Sebastian2014N,Medici2014PRL,Keimer2015N},
topological order~\cite{Jiang2012NP,Wen2017RMP,Wen2019S} or quantum
criticality~\cite{Hertz1976PRB,Senthil2004S,Sachdev2011PT},  often
requiring non-perturbative descriptions. 

A key difficulty in addressing QMBs arises from the exponential growth
of their Hilbert space with system size. This complexity is not only
combinatorial but is deeply connected to the structure of quantum
entanglement, governing how information is distributed in many-body wavefunction~\cite{Dalmonte2018NP,Lukin2019S,Cirac2021RMP,Joshi2023N,Zhao2025NP}.
Over the past decades, major advances have leveraged this structure in
different ways to solve
QMBs~\cite{Bauer2011JSMTE,LeBlanc2015PRX,Carleo2017S,Orus2019NRP}, for
example through tensor-network
methods~\cite{Stoudenmire2012ARCMP,Orus2019NRP,Mortier2025SP} addressing
low-entanglement regimes~\cite{White1992PRL,Peschel1999,Garcia2004PRL,Schollwock2005RMP,Schollwock2011AOP,Amaricci2008PRL,Di-Dio2015PRB,Chan2016JCP,Orignac2017PRB,Zhai2021JCP,Huang2021NCS,Sehlstedt2025A}. 
Many of these approaches ultimately rely on the repeated application of
the Hamiltonian to a quantum state, which forms the computational kernel
of many eigen-solvers and strategies to compute dynamical correlations. 
Yet, it remains unclear how this fundamental operation can be organized in a
physically transparent way coordinated with entanglement to
reach scalability in QMBs~\cite{Dolfen2006,Rincon2010CPC,Stoudenmire2013PR,Sharma2015CPC,Siro2016CPC,Chan2016JCP,Brabec2020JOCC,Levy2020SICHPCNSA,Zhai2021JCP,Huang2021NCS,Amaricci2022CPC,Menczer2024JCTC}.

In this work we show that the solution to this problem lies in
exploiting the entanglement structure as an organizing principle. An
arbitrary entanglement cut induces a bi-partition of the Hilbert space
and the decomposition of the (pure) wavefunction into
symmetry-resolved components. This bipartite representation provides a
natural way for distributing the many-body state across computing cores reflecting a decomposition of the entanglement spectrum. Consequently, scalability follows the intrinsic
structure of the quantum correlations rather than an externally imposed
layout.
The action of the Hamiltonian is reduced into a minimal set of local
operations combined with communication-optimal  data rearrangements
preserving locality~\cite{Dolfen2006,Amaricci2022CPC}. This framework
relies only on two general assumptions: (i) a $K$-local Hamiltonian and
(ii) the presence of a symmetry group. The resulting scheme
unlocks access to strong scalability in QMBs independently of the
high-level numerical method, establishing a direct link between
entanglement and computational performance. The same features that
control the entanglement
spectrum~\cite{Dalmonte2018NP,Goldstein2018PRL,Bonsignori2019JPMT,Zhao2025NP,Bellomia2026PR}
determine load balancing and communication cost, making scalability an
emergent property of the underlying quantum state. We demonstrate this
behavior using exact diagonalization (ED)~\cite{Crippa2025SPC} 
and density matrix renormalization group (DMRG) solutions of 
paradigmatic QMBs, showing asymptotic near-linear scaling for 
sufficiently large systems.

We consider $K$-local Hamiltonian systems of the form:
\begin{equation}
  H = \sum_{n=1}^{K} \sum_{i_1,\ldots,i_n} h^{(n)}_{i_1,\ldots,i_n} 
  \label{Hmodel}
\end{equation}
describing interacting particles on a finite lattice $\SS$ of size
$|\SS| = S$. Each term of the sum can be written as a product of local
operators, i.e., $h^{(n)}_{i_1,\ldots,i_n} \propto \prod_{m=1}^n
\OO_{i_m}$. We suppose that the local Hilbert-space $\HH_i$ has a finite
dimension $D$ and that it exists a symmetry group $\QQ$, i.e.
$[\QQ,H]=0$, which is not spontaneously broken by the ground-state or the
dynamics under consideration. The presence of $\QQ$ introduces a
factorization of the Hilbert-space $\HH$ and, consequently, of the
Hamiltonian $H$ into symmetry-blocks $H^\qa$ indexed by a set of quantum
numbers $\qa$. 

Further, we take into account an arbitrary entanglement cut (EC)
$\CC$ creating a bi-partition L and R such that $\HH=\HH_L\otimes\HH_R$.
$\CC$ is required to preserve the symmetry group $\QQ$. As a result, we
can associate to each partition a set of quantum numbers $\qa_{\a=L,R}$
such that $\qa_L \oplus \qa_R = \qa$, where $\oplus$ indicates the
composition of quantum numbers dictated by the commutative nature of
$\QQ$. 
This condition is satisfied by $N_Q$ tuples $q=(q_l,q_r)$ enumerating 
all possible combinations of quantum numbers $q_l$ and $q_r$, 
associated with the $L$ and $R$ partitions, respectively. 
In the following we assume $\QQ$ is Abelian. 
Examples of such partitions, including their application to
the symmetry group space, are shown in \figu{Fig1}(a-c).

The symmetry-block Hamiltonian takes the general form:
\begin{equation}
    H^\qa
     \!=\!\sum_{q=(q_l,q_r)} 
     \!\left[
     H^{q_l}_L \!\otimes\! \11_R^{q_r} 
     \!+\! 
     \11_L^{q_l}\!\otimes\! H^{q_r}_R 
     \!+\! 
     \sum_{m,k}L^{q_lk}_m \!\otimes\! R^{q_r k}_m
     \right]
  \label{Hmodel2}
\end{equation}
where $H^{q_l(q_r)}_{L(R)}$ are diagonal blocks of the Hamiltonian of the $L$
($R$) partition, while $L^{q_l k}_m$ ($R_m^{q_r k}$) are off-diagonal blocks 
of operators with support in $L$ ($R$), connecting a symmetry sector with 
quantum number $q_l$ ($q_r$) in $L$ ($R$) with another sector labeled by $k$. 
The index $m$ runs over the finite number of
coupling terms which connect the two partitions~\cite{Chan2016JCP,Zhai2021JCP}.  
In particularly symmetric cases this term reduces to a diagonal
form~\cite{Amaricci2022CPC}. The bi-partition imposes the following 
representation of the wavefunctions (where we simplify the notation referring only to the  index $q$):
\begin{equation}
  \label{PsiDecompQ}
  \ket{\Psi} = 
  \sum_{q} \sum_{l_q=1}^{d^L_q}\sum_{r_q=1}^{d^R_q}
  \Psi^{q}_{r_q l_q}\ket{{l_q}}\otimes \ket{{r_q}}
\end{equation}
where $\{\ket{l_q}\}$ and$\{ \ket{r_q}\}$ are orthonormal basis states
of the $L$ and $R$ partitions, respectively, of dimensions
$d^L_q$ and $d^R_q$, while $\Psi^{q}_{r_q l_q}\in\CCC$. This establishes a
natural isomorphism between the quantum states and the set of matrices
$\hat{\Psi}^{q}$, see \figu{Fig1}(d). 

We can now match the wavefunction distribution with the entanglement 
structure encoded by the block matrices $\hat{\Psi}^{q}$. 
Specifically, as each term $\hat{\Psi}^{q}$ 
is defined on the symmetry-resolved block basis $\ket{l_q}\otimes\ket{r_q}$, for
each state $\ket{r_q}$ in the $R$ partition, the set of states
$\ket{l_q}$ in the $L$ partition can be distributed across $P$
processes. This corresponds to a column-wise distribution of each block matrix 
$\hat{\Psi}^{q}$ so that the process $p=1,\dots,P$ is assigned with
$\Theta_{q} = [d^L_q/P]$ columns, see \figu{Fig1}(d).

\begin{figure}
  \centering
  \includegraphics[width=1\linewidth]{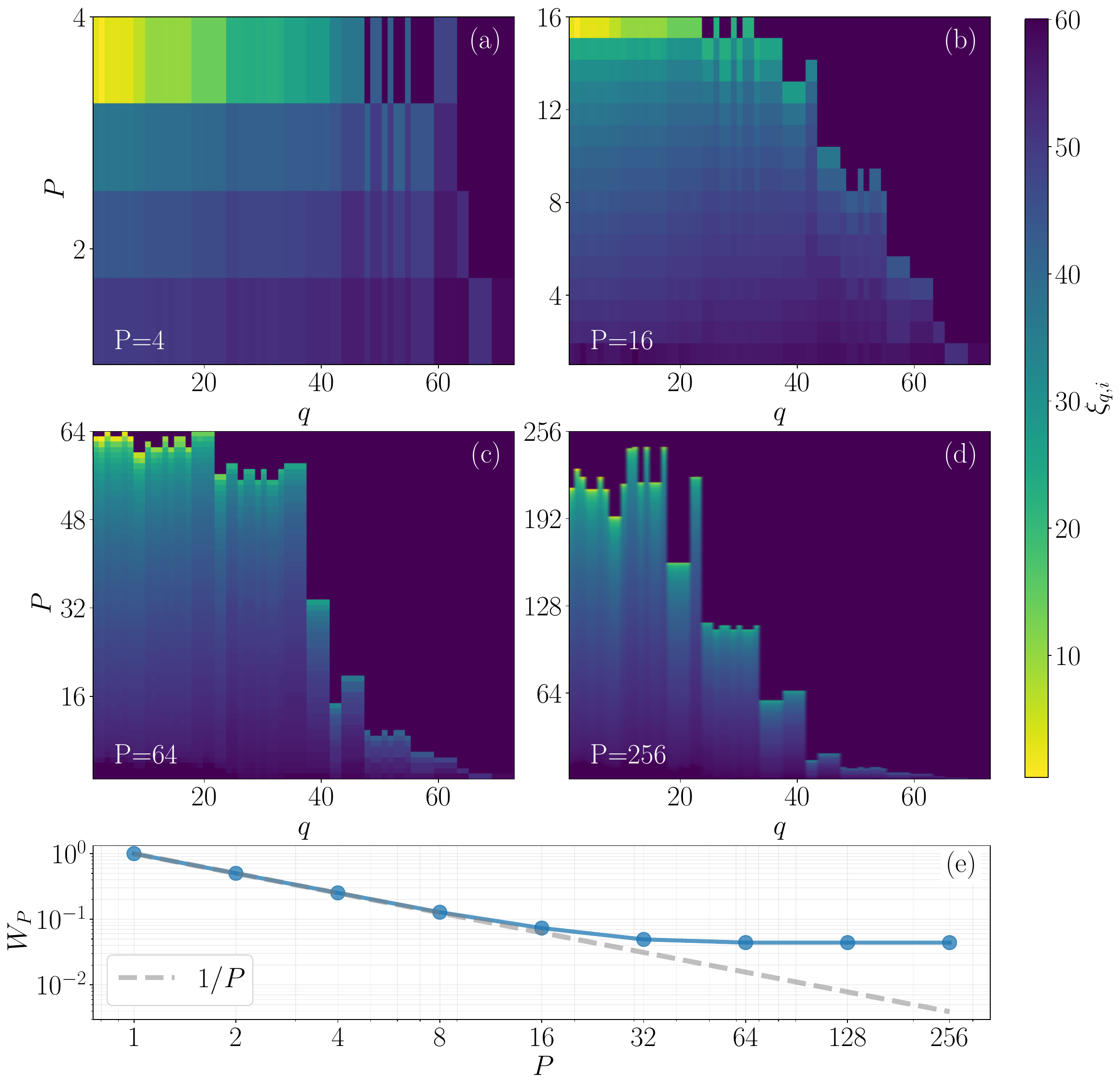}
  \caption{\textbf{Distributed entanglement spectrum.} 
  Evolution of the distribution of the spectrum $\xi_{q,i}$ of the entanglement Hamiltonian
  $H_E=-\log\rho_L$ as a function of the symmetry-sectors label $q$ and the
  number of processes $P$ (a-d). 
  Data are from DMRG solution of the Hubbard
  model with $U/t=2$, $S=20$ and total bond dimension $\chi=4000$. 
  (e) Inverse participation ratio $W_P$ as a function of $P$ for the same
  system.}
  \label{figEntang}
\end{figure}
In this setup the scalability of the problem is controlled by the
structure of the entanglement spectrum. The parallel decomposition
can be interpreted as an entanglement-resolved representation of the
quantum state. This is illustrated in \figu{figEntang}, where we analyze
the  spectrum of the entanglement
Hamiltonian~\cite{Dalmonte2018NP,Giudici2018PR}
$H_E=-\log\rho_L=-\log\Tr_R \ket{\Psi_0}\bra{\Psi_0}$, with
$\ket{\Psi_0}$ the ground-state, for increasing $P$.  
Each process carries a subset of the Schmidt components so that the
distributed wavefunction corresponds directly to a partition of the
entanglement spectrum $\xi_{q,i}$ and thus define the intrinsic
structure of the distributed data. Increasing $P$ progressively resolves
the entanglement weight across the symmetry-sectors, leading to a more
uniform distribution.
To quantify this redistribution, we report in \figu{figEntang}(f) the
process-resolved inverse participation ratio $W_P= \sum_{q \in
\mathcal{Q}_p} W^2_q$ of the entanglement weight
distribution, where $\mathcal{Q}_p$ is the set of symmetry-sectors 
assigned to process $p$ and $W_q = {\sum_i e^{-\xi_{q,i}}}/{\sum_{q,i} e^{-\xi_{q,i}}}$. 
$W_P$ remains largely sub-extensive over the number of
processes, indicating partial but non-trivial scrambling of the
entanglement spectrum. This determines how efficiently the wavefunction
can be distributed and therefore control scalability and the
communication cost, see Figs.~(\ref{figScaling})-(\ref{figSymFrag}).

\begin{figure*}
  \centering
  \includegraphics[width=1\linewidth]{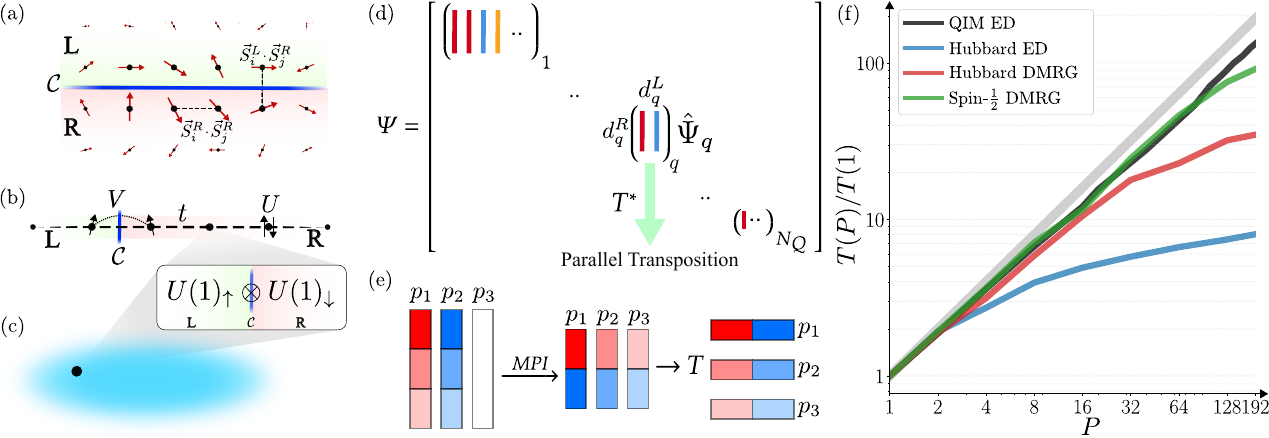}
  \caption{
  \textbf{Setup and scaling of the entanglement-informed wavefunction distribution.}
  (a-c) Examples of system bi-partition. The blue line indicates
  the EC $\CC$ separating the system in $L$ and $R$ partitions for a 2d
  spin model (a), a 1d Hubbard model with hopping $t$, local $U$
  and non-local interaction $V$ (b) and a quantum impurity model (c) featuring a cut in the
  spin symmetry group space. (d) Representation of the distributed wavefunction. Vertical colored segments corresponds to the $L$ states
  distribution of the wavefunctions blocks $\hat{\Psi}^{q}$. (e)
  Schematic description of the parallel transposition  $T^*$ using $P=3$
  and $d_q^L=2$. (f) Scaling of the speed-up $T(P)/T(1)$ in
  different models, methods and ECs. The data are obtained with ED and DMRG for symmetry-sectors of total 
  dimensions $D\simeq 10^7$. The DMRG data are obtained for $S=20$ chains,
  $\chi=4000$ and $\CC$  placed at the center of the lattice. The ED
  solution of the 1d Hubbard model is for $S=14$ and $\CC$ as in panel
  (b). The impurity model is solved using ED for $S=15$ and $\CC$ as in panel (c).
  }
  \label{Fig1}
\end{figure*}
%
The entanglement-resolved representation yields a substantial
simplification of the Hamiltonian application $H^{\qa}\ket{\Psi}$,
which constitutes the core operation in any Krylov-based method for
solving QMBs.
The operators residing in the $R$ partition can be applied locally,
however contractions with operators in $L$ require a costly data
exchange to reconstruct the corresponding basis states, spoiling any
potential advantage. The challenge therefore shifts to preserving data
locality so that high-throughput operations can be performed using only
process-local data. A straightforward solution is to transpose the blocks
$\hat{\Psi}^{q}$ which is however an expensive collective operation.
Fortunately, the distributed-memory parallel framework makes available a
procedure transferring data so that the $m$-batch owned by process $p$ is
sent out to process $m$ and placed as batch $p$, see
\figu{Fig1}(d-e)~\cite{MPI2,Dolfen2006,Amaricci2022CPC}. This implements
a parallel transposition,  denoted by ${T^*}$, that minimize the amount
of data transfer. Use of $T^*$ removes the communication congestion
and unlocks optimal scaling. 

The Hamiltonian application essentially corresponds to perform three local
and scalable operations. The product of the term $\11^{q_l}_L
\otimes H^{q_r}_R$ involves access to $R$ states and so it does not
require any data transfer, making this operation local on the process
and embarrassingly parallel.
On the contrary, the product of $H^{q_l}_L \otimes \11^{q_l}_R$
requires reconstructing states of the $L$ partition. This is readily
achieved performing a $T^*-$transposition of the block $\hat{\Psi}^{q}$
before acting with the operator $H^{q_l}_L$ and, finally, to
$T^*-$transpose the result back in order to reconstruct the contribution
to the distributed wavefunction.
The action of the term $L^{q_l k}_m \otimes R^{q_r k}_m$, 
for any $m$ and $k$, 
appearing in \equ{Hmodel2} requires some algebraic manipulations.
First, the product is reformulated as $ \left[ L^{q_l k}_m \otimes
R^{q_r k}_m \right]{\rm Vec}(\hat{\Psi}) = {\rm Vec}\left[ R^{q_r k}_m \cdot
\hat{\Psi}^{k} \cdot {L^{q_l k}_m}^T\right] $ where ${\rm Vec}(A)$ transforms
the $a\times b$ matrix $A$ into a vector of length $ab$ by stacking its
columns. The evaluation proceeds computing the intermediate term
$R^{q_r k}_m\cdot\hat{\Psi}^k=C^{q_r k}_m$. Since $\hat{\Psi}^k$ is distributed
according to the $R$ states, this operation is local in the memory and
embarrassingly parallel. The auxiliary matrix $C^{q_r k}_m$ inherits the
same distribution as the wavefunction, a key property that unlocks the
scalable computation of the remaining term. Finally, the product  is recast as
$\left[L^{q_l k}_m \cdot {C^{q_r k}_m}^{T^*} \right]^{T^*}$ so it can be completed using only two additional $T^*$ operations.
Thus, leveraging entanglement-informed wavefunction distribution it is
possible to reduce the application of the Hamiltonian to a set of
massively parallel local operations and few optimized global data
exchange, removing the intrinsic bottleneck of any Krylov-based QMBs
simulation.

This setup is model-independent and can be applied to any QMBs
fulfilling the initial assumptions. To assess this generality we compare, in
\figu{Fig1}(f), the scaling behavior for several benchmarks, including ED
and DMRG solutions of the 1d Spin-$\tfrac{1}{2}$ Heisenberg (SM),
Hubbard  (HM) and Quantum Impurity model (QIM).  
The latter features an EC coinciding with the spin symmetry
group factorization, see \figu{Fig1}(c), corresponding to the wavefunction 
being distributed as a product of two independent spin blocks. Notwithstanding the
differences in the nature of the problems all the reported solutions
show an optimal scaling.


In the following, we analyze the scaling behavior using DMRG as a tunable
testbed to probe further properties of this representation. 
The total size is controlled by the bond dimension $\chi$ while 
the sizes of the symmetry-sectors on each partition get truncated 
to their symmetry-blocks bond dimensions $\chi_q$, i.e. $d^L_q\rightarrow \chi^L_q$ 
and $d^R_q\rightarrow \chi^R_q$ with  $\sum_q \chi^{\alpha=L,R}_q=\chi$ and 
$\chi_{q}=\chi^L_q\chi^R_q$. 
In \figu{figScaling}, we report the speed-up $T(P)/T(1)$ for the 1d SM, where $T(p)$ is the Hamiltonian application time with $p$ processes. 
The scaling improves with increasing $\chi$, 
with all curves displaying strong-scaling behavior saturating at larger $P$. 
The result for the largest  $\chi$ demonstrates the asymptotic nature of the 
scaling, displaying a near-linear behavior $T(P)\sim 1/P$~\cite{Dolfen2006,Brabec2020JOCC,Amaricci2022CPC}. 
The residual deviation is described by a modified Amdahl's law,
indicating that the non-parallelizable fraction remains small even at
large process count~\cite{Gustafson1988C,Gustafson2011}. This confirms
that multiple use of $T^*$ introduces only a weak sub-leading correction.
The practical limitation arises from the finite size of the local data
share, as shown in \figu{figScaling}. Increasing $P$ reduces the
parallel efficiency and the amount of data stored per CPU, until the
latter becomes comparable to the 
local cache memory size. Beyond this point,
latency effects dominate over raw floating-point throughput. This
asymptotic scaling is therefore not set by the algorithmic structure,
but by the hardware-imposed locality of the distributed wavefunction.

\begin{figure}[t!]
  \centering
  \includegraphics[width=1\linewidth]{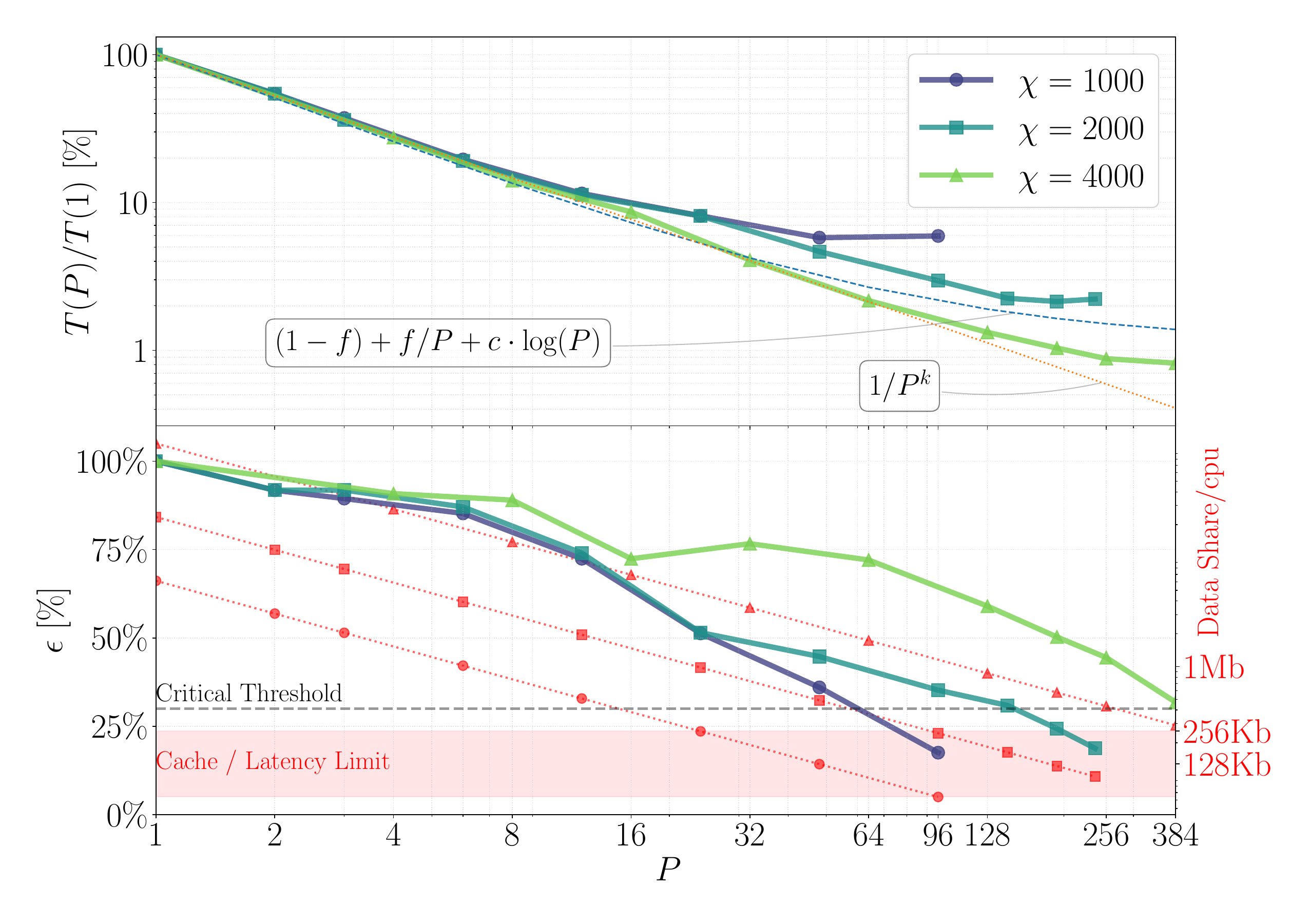}
  \caption{\textbf{Asymptotic scaling of the QMBs solution.}
    \textit{Top panel}: parallel speed-up $T(P)/T(1)$ as a function
  of the number of processes $P$. Data are from the DMRG solution of the
  1d SM with $S=20$. The dashed line indicates the Amdahl's law fit
  ($f=0.95$). The dotted line indicates power-law fit ($k=0.93$). {\it
  Bottom panel}: parallel efficiency $\epsilon=\tfrac{T(1)}{P T(P)}$ (left $y$-axis) and data
  share per CPU (right $y$-axis) as a function of $P$. Data as in the
  top panel. The marked region indicates the latency limit regime. The
  dashed line indicates the corresponding efficiency critical threshold.
  }
  \label{figScaling}
\end{figure}

\begin{figure}
  \includegraphics[width=1\linewidth]{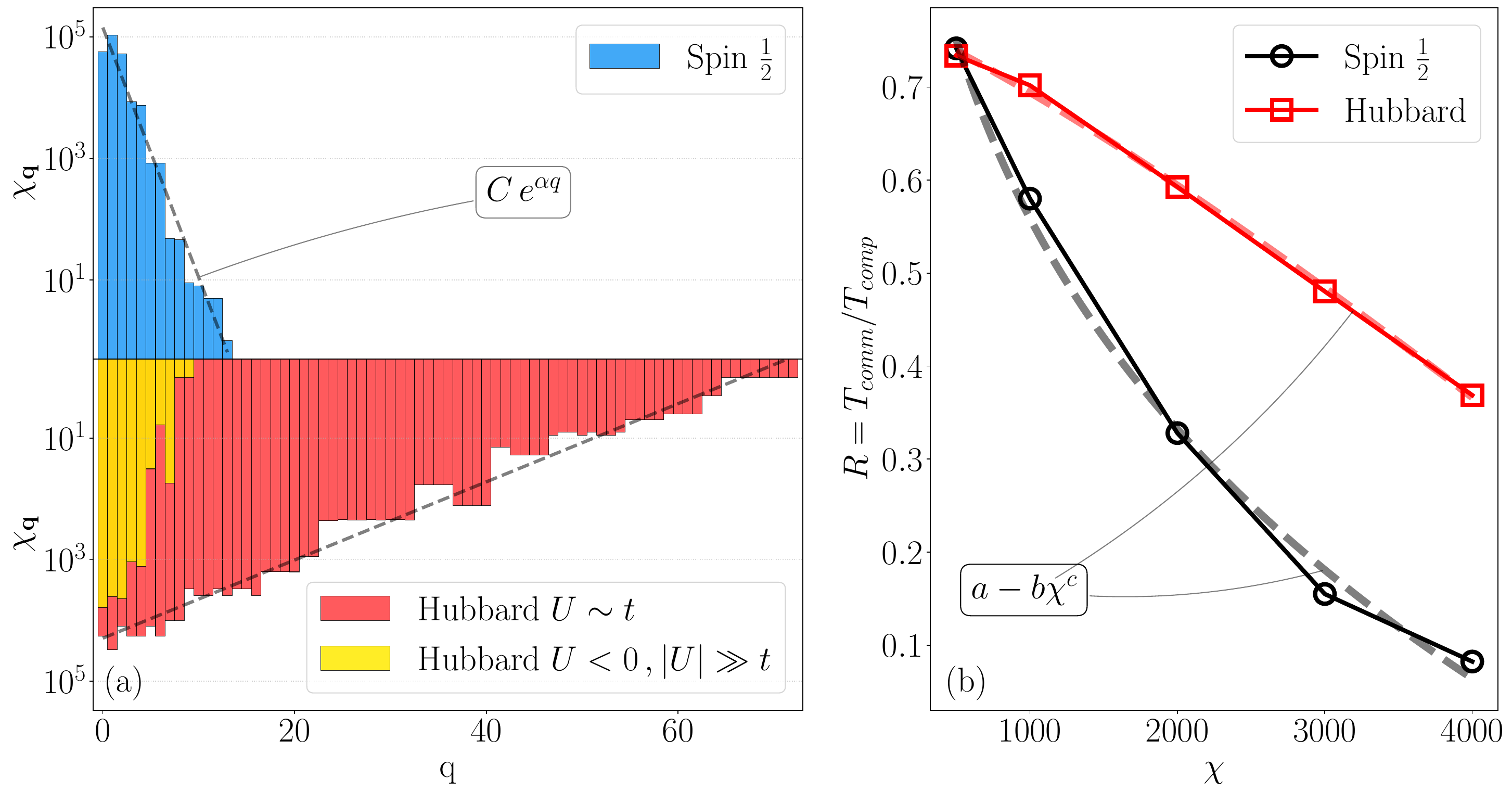}
  \caption{ \textbf{Symmetry fragmentation and communication overhead.}
  (a) Wavefunction symmetry-blocks bond dimensions $\chi_q$
  distribution as a function of the sector index $q$. Data from DMRG
  solution with $S=20$ sites, $P=64$,
  $\chi=4000$ of the SM $J=1$ (top), the HM $U=2t$ and attractive HM $|U|>>t$ (bottom).  
  The dashed lines are fit  with exponential distributions:  $\a=0.95$ (SM), $0.15$ (HM) and
  $1.11$ (attractive HM). 
  (b) Communication-over-computation time ratio
  $R=T_{comm}/T_{comp}$ as a function of the total bond dimension
  $\chi$. Data as in panel (a). The dashed line indicates the fit with
  effective model with $[a,b,c]=[1.53,0.45\,10^{-2},0.17]$
  (SM) $[0.77,0.71\,10^{-5},1.1]$ (HM). 
  }
  \label{figSymFrag}
\end{figure}
While $\chi$ sets the overall entanglement and computational cost,  
its distribution across the symmetry-sectors determines how this cost is partitioned and communicated. This is ultimately determined by the
fragmentation of the wavefunction induced by the EC. The
\figu{figSymFrag}(a) shows the distribution of the symmetry-sectors bond
dimensions $\chi_q$ for spin and fermionic problems. 
In all cases the
decay follows an overall exponential law $\chi_q =C e^{-\alpha q}$, with
a rate that becomes substantially slower as the symmetry group
increases. 
The attractive HM in the strong-coupling limit provides an
instructive interpolation: once holon/doublon formation suppresses the
independent spin sectors, the distribution collapses back to the same
steep decay of the spin case.

Such {\it symmetry fragmentation} sets the
communication-over-computation ratio $R=T_{comm}/T_{comp}$, see
\figu{figSymFrag}(b). While the overall computational cost roughly
scales as $O(\chi^3)$, i.e. total volume of the distributed blocks,
communication is controlled by their exposed \textit{ surface} under the
parallel transposition. 
Although $\chi_q$ decays exponentially, the complementary cumulative
distribution displays a broad approximately power-law behavior over a 
significant range~\cite{SeeEndMatter}. A few large sectors
coexist with a proliferation of many smaller ones, resulting in an
effectively heavy-tailed distribution of block sizes. This produces a
correction to the usual surface-to-volume argument, leading to a slower
decay of communication cost~\cite{SeeEndMatter}. The fitted form captures
this behavior accurately: spin systems approach the compute-dominated
regime much faster, whereas the electrons retains some communication
overhead.

To conclude, we have introduced a general framework 
applicable to most state-of-the-art strategies for QMBs simulations 
in which the entanglement structure of the wavefunction determines its
optimal parallel representation.  An arbitrary EC maps the
quantum state distribution onto a partition of the entanglement spectrum, 
reducing the Hamiltonian application to local contractions and 
communication-optimal operations. 
A near-linear asymptotic scaling follows directly from the
state’s entanglement structure, while the
heavy-tailed distribution of symmetry-sectors sets an irreducible
communication cost. 
%
%
This framework provides a unified, method-independent route to scalable 
QMBs simulations by directly linking entanglement to computational performance. 
Assuming only locality and symmetry properties, the construction serves as a low-level framework  applicable across diverse many-body approaches.
More broadly, entanglement
emerges not only as a measure of quantum complexity, but as a resource
to organize and scale up classical QMBs simulations, enabling
entanglement-informed strategies for large-scale computations.

\paragraph*{Acknowledgments.} 
We are grateful to C.~Mejuto-Zaera, M.Fabrizio and M.Collura for many insightful discussions and a critical reading of the manuscript. We also thank M.Dalmonte, D.J.Garc\'ia and S.de Palo. We acknowledge financial support from the National Recovery and Resilience Plan PNRR MUR Project No. PE0000023-NQSTI.

\bibliography{references}

\section*{End Matter}\label{EndMatterSec} 
\begin{figure}
  \centering
\includegraphics[width=1\linewidth]{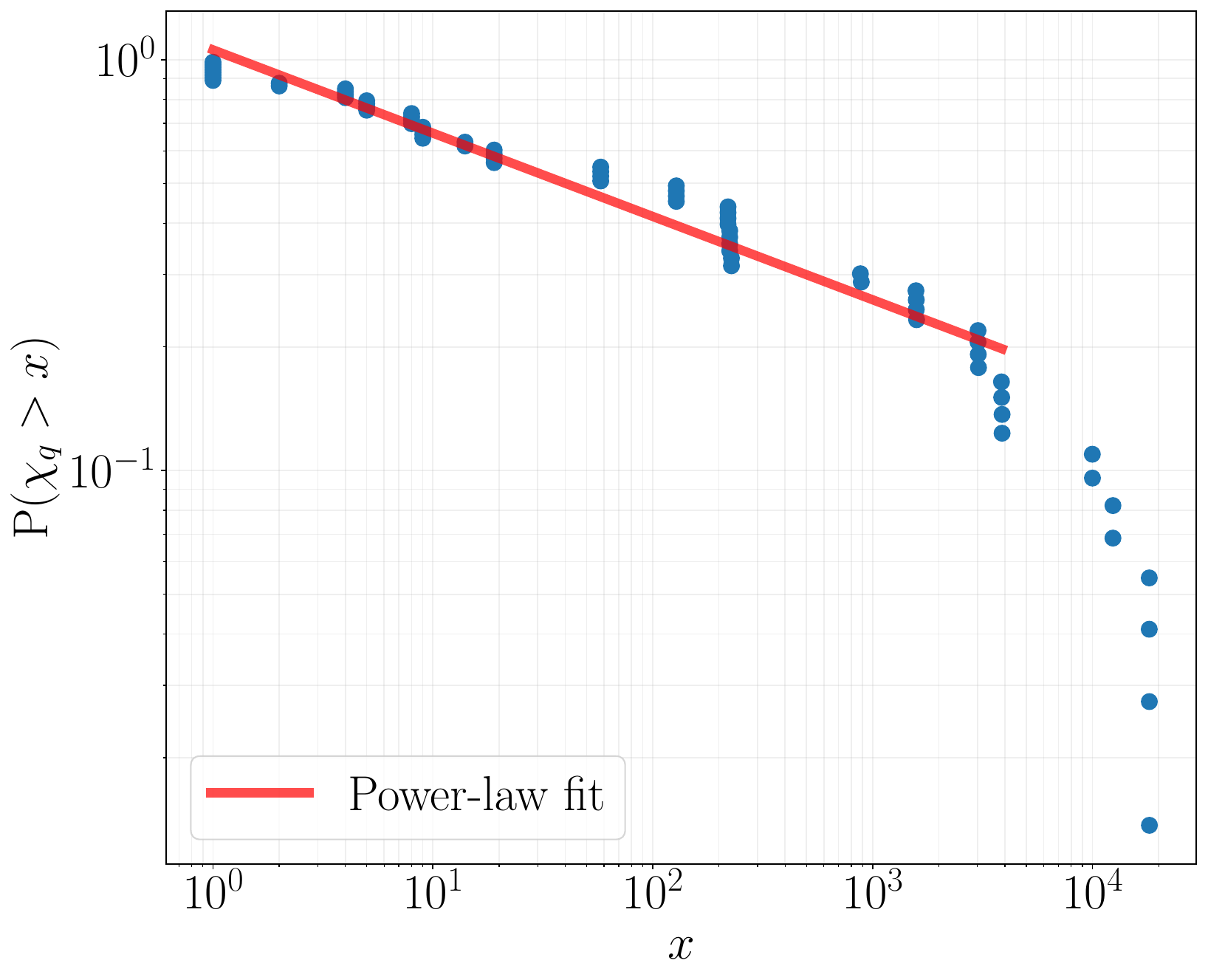}
  \caption{Complementary cumulative distribution function of the
  symmetry-sectors bond dimension $\chi_q$ for the Hubbard model. Data
  are from DMRG solutions with $S=20$, $P=64$, $U/t=2$ and $\chi=4000$.
  The red line is a power law fit, $\gamma=0.20$. }
  \label{figHeavyTail}
\end{figure}
\paragraph*{Effective description of symmetry fragmentation.}
The distribution of symmetry-sector  dimensions plays a central
role in determining computational load balancing and the associated
communication overhead (see \figu{figSymFrag}).
Here we focus on the specific case of symmetry-blocks bond dimensions $\chi_q$ from the DMRG solution. 

The overall decay
follows a global exponential-law $\chi_q= \chi^m e^{-\alpha q}$ where we expressed the constant $C$ in terms of the total bond dimension. Yet, a more refined analysis reveals that for systems with larger symmetry group
(e.g. the Hubbard model), the distribution exhibits an intermediate
regime characterized by heavy-tailed behavior. This feature is most
clearly captured by the complementary cumulative distribution function
(CCDF), defined as the probability $P(\chi_q > x)$ of observing a
symmetry-sector bond dimension larger than $x$.  

As shown in \figu{figHeavyTail}, the CCDF for the Hubbard model
displays an approximate linear behavior on a log-log scale over roughly three orders of magnitude. In this window, the distribution is
well described by a power-law 
$$
P(\chi_q > x) \sim A x^{-\gamma}\,.
$$

Such heavy-tail behavior indicates that, while a few large sectors
dominate the total memory, there is a significant proliferation of
intermediate-to-small sectors that overly contribute to the {\it exposed
surface} of the wavefunction during $T^*$-transpositions. The deviation
from the power-law regime around $\chi_q \sim 10^3$ signals the onset of
the dominant sectors that ultimately control the total computational
cost. 

To quantify the degree of fragmentation,  we introduce the effective
number of sectors $N_{\text{eff}}$ via the inverse participation ratio
(IPR) of the normalized symmetry-block dimensions 
$w_q = \chi_q /\sum_{q'} \chi_{q'}$. 
The effective participation ratio 
$$
N_{\text{eff}} = 1/\sum_q w_q^2
$$ 
measures the degree of  non-uniformity of the distribution and provide
an estimate of the number of dominant sectors contributing to the total
computational load. Results for the Hubbard model shows that
$$
N_{\text{eff}} \ll N_Q\,,
$$ 
consistent with a heavy-tailed
distribution in which a relatively small subset of the symmetry-sectors carries most of the total dimension.

Based on these observations, we derive a simple effective description of
the communication-over-computation $R$. We assume that: (i) the
dimensions of the symmetry-blocks follow $\chi_q=\chi^m e^{-\alpha q}$
and (ii) there exists a crossover index $q^*$ identifying the
$N_{\text{eff}}$ dominant sectors. By definition, we obtain 
$$
q^*
= \frac{m}{\alpha}\ln{(\frac{\chi}{\Pi})}\,,
$$
where $\Pi^m=\chi_{q^*}$.  

Assuming that the computational cost scales as the cube of the sector
dimensions and is dominated by the largest blocks, we estimate:
$$
T_{cpu} \simeq \frac{1}{q^*}\int_0^{q^*}\chi_q^3 dq 
\simeq\frac{\Pi}{3m}\chi^{3m-1} \left[1+O\left(\frac{\Pi}{\chi}\right)\right]
$$
where we considered that $\chi^* \ll \chi$ and expanded
$\ln{(\frac{\chi}{\Pi})}$  to the linear order. 

Similarly, we decompose the communication cost into contributions from large and small sectors, $T_{comm}=T_{comm}^> + T_{comm}^<$. 
The first term, $T_{comm}^>$, corresponds to the dominant sectors and is assumed to scale weakly with $\chi$. 
The second term, $T_{comm}^<$, accounts for the proliferation of intermediate-to-small sectors. This contribution must be evaluated relative to the full distribution, since the dominant sectors  progressively deplete the available weight as $\chi$ increase. We modeled this contribution through an effective (fractal) exponent $D$:  
$$
T^<_{comm} = \int_{q^*}^{\infty}[\chi_q]^D dq \simeq \Lambda - \frac{\Pi^D}{\alpha D}\chi^{(m-1)D}
$$
where $\Lambda$ expresses the total cost. By combining these estimates, the communication-over-computation ratio $R$ takes the effective form:
$$
R=T_{comm}/T_{cpu}\simeq a - b \chi^{c}
$$
with $c=1-D+m(D-3)$, consistently with the scaling observed in \figu{figSymFrag}(b).

\end{document}